Handreichung für Lehrpersonen:

# Die Stromversorgung der ISS

## Klassen 7/8

Markus Nielbock

22. Januar 2019

## Zusammenfassung

Diese Aktivität ermöglicht den Schülerinnen und Schülern, die Stromversorgung der Internationalen Raumstation zu untersuchen. Falls vorhanden, ermitteln sie die augenblicklichen Parameter des elektrischen Systems aus der Telemetrie der ISS in Echtzeit. Ansonsten können sie Archivdaten nutzen, die den Arbeitsunterlagen beigefügt sind. Hieraus berechnen sie die von den Solarzellen zur Verfügung gestellte elektrische Leistung.

## Lernziele

Die Schülerinnen und Schüler

- berechnen die Fläche der Sonnensegel der ISS,
- berechnen die elektrische Leistung,
- berechnen den Anteil der Strahlungsintensität der Sonne, die in elektrische Leistung umgesetzt wird,
- berücksichtigen bei den Berechnungen den Wirkungsgrad einer Solarzelle.

## Materialien

- Arbeitsblätter (erhältlich unter: http://www.haus-der-astronomie.de/raum-fuer-bildung)
- Stift
- Taschenrechner
- Computer/Tablet/Smartphone mit Internetzugang (optional)

## Stichworte

Raumstation, ISS, Solarzellen, Solarkonstante, Elektrizität, Spannung, Stromstärke, Leistung

## Dauer

90 Minuten

Unterrichtsmaterialien zur ISS-Horizons-Mission von Alexander Gerst

In Kooperation mit DLR

## Hintergrund

### Die Internationale Raumstation

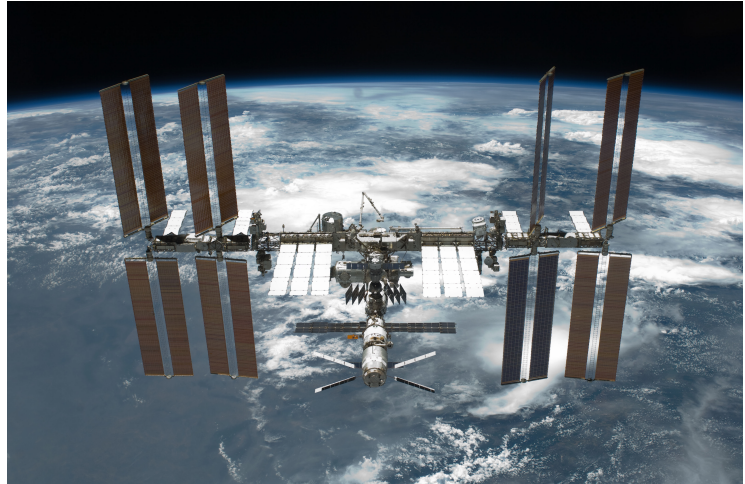

**Abbildung 1:** Die ISS im Jahre 2011 (Bild: NASA).

Seit 1998 wird die Internationale Raumstation (ISS, Abb. 1) aufgebaut (Loff 2015) und mittels einzelner Module (Abb. 2) ständig erweitert (Zak 2017). Ihr Betrieb ist bis mindestens 2024 vorgesehen, wahrscheinlich aber sogar bis 2028 möglich (Sputnik 2016; Ulmer 2015). Die gesamte Struktur hat eine Masse von 420 t. Sie ist 109 m lang, 73 m breit (Garcia 2018b) und 45 m hoch (ESA 2014). Auf einer Bahnhöhe von etwa 400 km benötigt die ISS für eine Erdumrundung ungefähr 92 Minuten (Howell 2018).

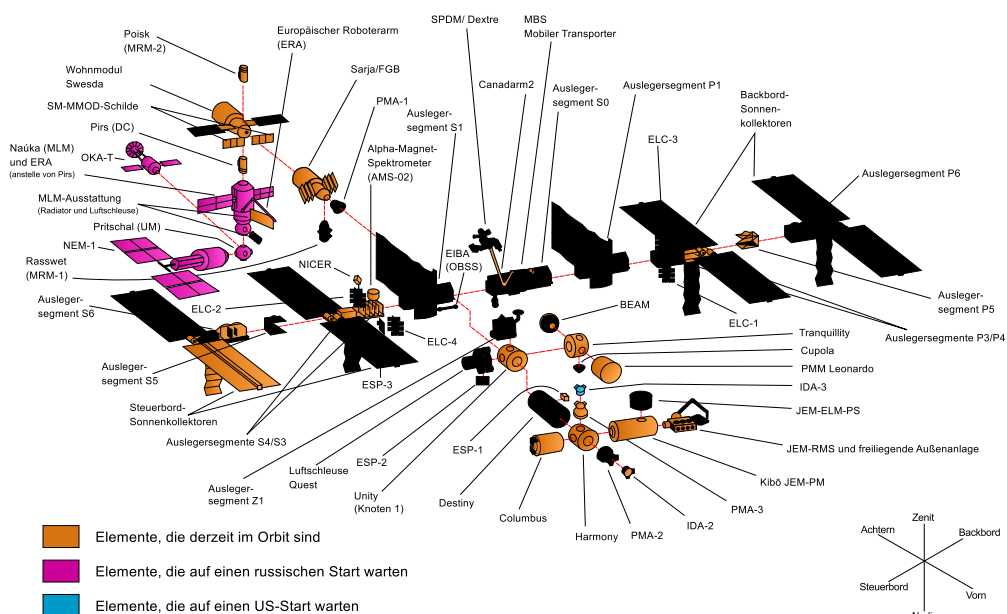

**Abbildung 2:** Die Module der ISS im Juni 2017 (Bild: NASA).

Die ISS ist ein internationales Projekt mit derzeit 15 beteiligten Nationen (ESA 2013; Garcia 2018a). Sie dient als wissenschaftliches Forschungslabor für Fragestellungen, in denen z. B. der Einfluss der Gravitation auf der Erde bei Experimenten hinderlich ist. Es werden aber auch medizinische Themen der Astronautik behandelt, um langfristige Missionen innerhalb des Sonnensystems vorzubereiten.

In Abb. 2 erkennt man neben den verschiedenen Modulen auch die Sonnenkollektoren, die die ISS mit Strom versorgen. Diese Übung befasst sich mit der Stromversorgung der ISS. Daher werden nachfolgend einige Grundlagen der Elektrizität erläutert, die bei der Lösung der Aufgaben hilfreich sind.

**Grundbegriffe der Elektrizität**

Der Elektromagnetismus ist eine der vier Grundkräfte der Physik. Er setzt sich aus der Elektrizität und dem Magnetismus zusammen. Der Grundbaustein der Elektrizität ist die elektrische Ladung, die entweder positiv oder negativ sein kann. Gleichnamige Ladungen stoßen sich ab. Unterschiedliche Ladungen ziehen sich an. Die Ladung eines Elektrons $e$ wird hierbei als Elementarladung angesehen. Sie ist eine Naturkonstante. Ihre Einheit ist das Coulomb C, benannt nach dem französischen Physiker Charles Augustin de Coulomb.

$$e = 1{,}602 \cdot 10^{-19} \text{ C}$$

Das Formelzeichen für die Ladung ist $Q$. Pole mit unterschiedlichen Ladungen, haben das Bestreben, ein Gleichgewicht herzustellen, indem sie Ladungen austauschen. Wenn ein elektrischer Strom fließt, wandern Ladungsträger von einem Pol zum anderen. Das Vermögen, durch den Austausch von Ladungen elektrischen Strom aufrecht zu erhalten, nennt man die elektrische Spannung $U$. So sind in einer Batterie Ladungsträger getrennt angeordnet, so dass sie eine Spannung erzeugen. Innerhalb der Batterie kann kein Ausgleich stattfinden. Verbindet man die beiden externen elektrischen Pole mit einem elektrischen Leiter, bewegen sich die Ladungen. Es fließt ein Strom. Der elektrische Strom kann wegen der Bewegung der Ladungsträger somit als zeitliche Veränderung der Ladung interpretiert werden. Man bezeichnet diese Variation als Stromstärke, die mit $I$ bezeichnet wird. Die Stärke eines Stroms hängt davon ab, wie viele Ladungsträger transportiert werden und wie schnell sie wandern. Die Einheit ist das Ampere A. Fließen Ladungen von 1 Coulomb pro Sekunde durch den Stromleiter, entspricht das einer Stromstärke von 1 Ampere, benannt nach dem französischen Physiker André-Marie Ampère.

$$I = \frac{\Delta Q}{\Delta t}$$

Schaltet man einen Verbraucher dazwischen, z. B. eine Lampe oder einen Motor, verrichtet der elektrische Strom Arbeit. Dabei entnimmt jeder Verbraucher dem elektrischen Strom Energie. Das wirkt sich wie ein Widerstand $R$ aus, der bei gegebener Spannung $U$ die Stromstärke $I$ beeinflusst.

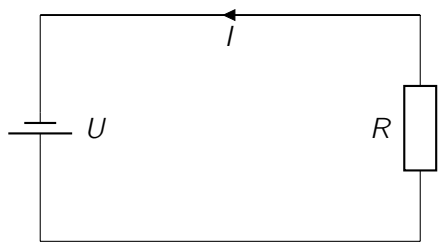

$$U = R \cdot I$$

Der Widerstand ist dabei eine Eigenschaft des Verbrauchers und hat die Einheit Ohm ($\Omega$), benannt nach dem deutschen Physiker Georg Simon Ohm. Bei einer Glühbirne hängt der Widerstand vom Material sowie der Länge und der Dicke des Drahts ab. Die Einheit der Spannung ist das Volt V, benannt nach dem italienischen Physiker Alessandro Volta. Die dabei verrichtete Arbeit entspricht einer elektrischen Leistung $P$, die von der am Verbraucher anliegenden Spannung als auch von der Stromstärke abhängt. Ihre Einheit ist wie überall in der Physik das Watt (W), benannt nach dem schottischen Wissenschaftler und Ingenieur James Watt.

$$P = U \cdot I = U \cdot \frac{U}{R} = \frac{U^2}{R}$$

Somit erzeugt ein Motor, der bei einer Spannung von 1 V eine Stromstärke von 1 A erfährt, eine elektrische Leistung von 1 W.

### Leuchtkraft der Sonne und die Solarkonstante

Die Sonne ist wie alle Sterne ein heißer Gasball. An ihrer Oberfläche beträgt die Temperatur 5778 K bzw. 6051 °C.

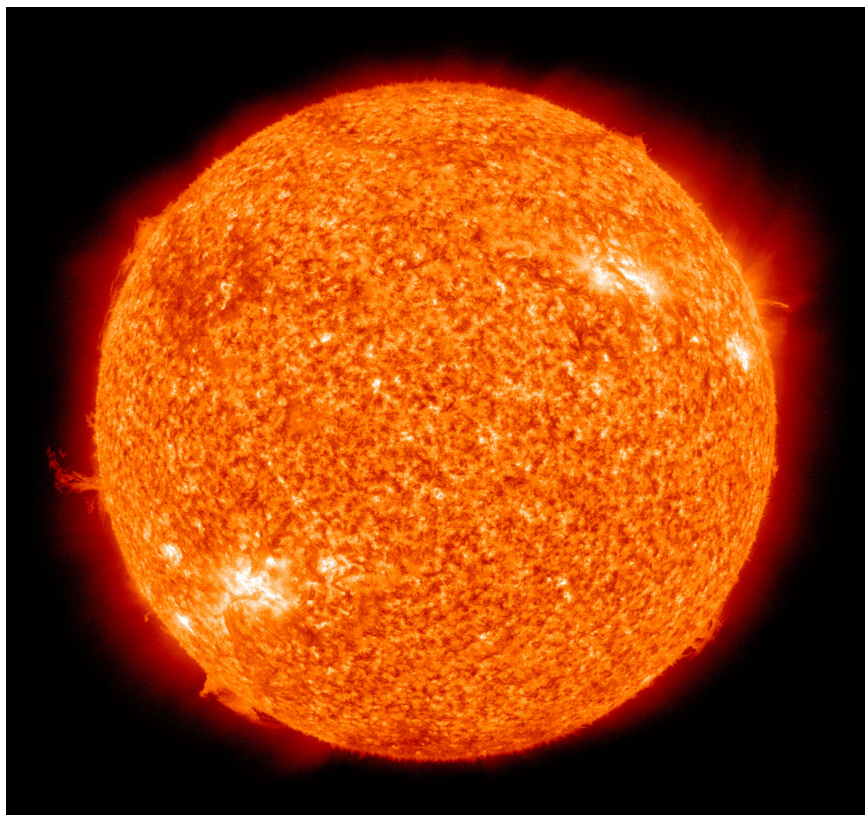

**Abbildung 3:** Eine Aufnahme der Sonne vom Solar Dynamics Observatory, einer Sonde im Weltall, die ständig die Sonne beobachtet (Bild: NASA/SDO (AIA)).

Dabei gibt sie ständig Strahlung mit einer bestimmten Leistung ab, also Energie pro Zeiteinheit. Bei Sternen nennt man diese Größe *Leuchtkraft L*. Die Einheit ist wie bei der üblichen Leistung das Watt.

$$L_\odot = 3{,}845 \cdot 10^{26}\,\text{W}$$

Gegenüber den hellsten Sternen im Universum ist das aber relativ wenig. Wir können uns mal genauer ansehen, was diese Leuchtkraft bezogen auf die Oberfläche der Sonne bedeutet. Ihr Radius beträgt $R_\odot = 695\,508$ km. Die Strahlungsleistung pro Flächenelement wird als Intensität $\mathcal{I}$ bezeichnet. Die Oberfläche der Sonne kann über die Formel für die Kugeloberfläche $O = 4 \cdot \pi \cdot r^2$ berechnet werden.

$$\mathcal{I}_\odot = \frac{L_\odot}{O_\odot} = \frac{L_\odot}{4 \cdot \pi \cdot R_\odot^2}$$

Umgerechnet sind dies 6300 W/cm². Die Gleichung zeigt auch, dass die gemessene Intensität vom Radius der betrachteten Kugel abhängt. Da für diese Übungseinheit wichtig ist, wie groß die Intensität am Ort der Erde ist, müssen wir eine Kugel betrachten, deren Radius dem Abstand der Erde von der Sonne entspricht. Dieser beträgt im Mittel $1,496 \cdot 10^{11}$ m.

Für die Intensität am Ort der Erde findet man dann:

$$\mathcal{I}_{\text{Erde}} = 1367 \, \frac{\text{W}}{\text{m}^2}$$

Dies ist die sogenannte Solarkonstante ($E_0$). Sie gilt streng nur für den mittleren Abstand zwischen Sonne und Erde. Wegen der leicht exzentrischen Erdbahn (Abb. 4) variiert der tatsächliche Wert der Intensität der Sonne im Laufe eines Jahres zwischen 1325 und 1420 W/m².

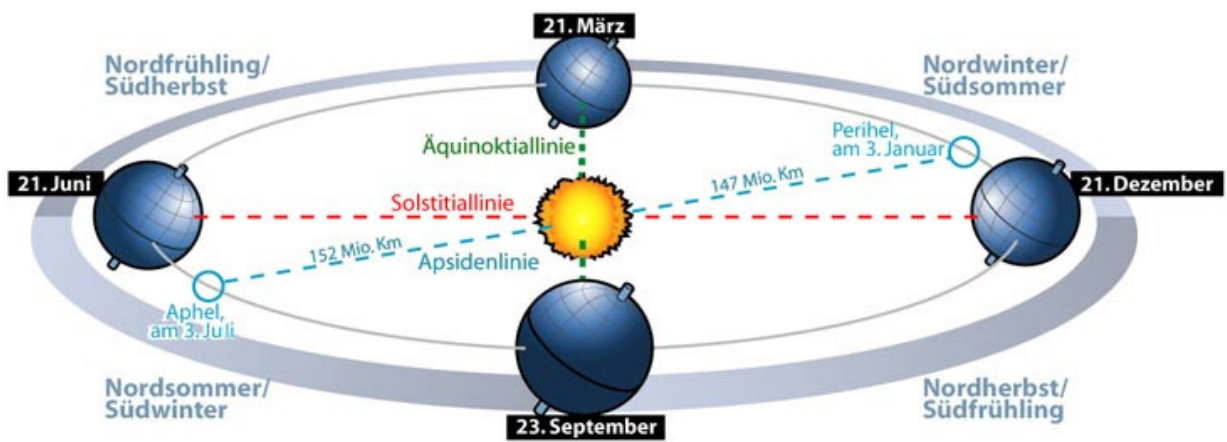

**Abbildung 4:** Darstellung der Bahn der Erde um die Sonne (Bild: Horst Frank (https://commons.wikimedia.org/wiki/File:Jahreszeiten99_DE2.jpg), https://creativecommons.org/licenses/by-sa/3.0/legalcode).

## Stromversorgung auf der ISS

Die Stromversorgung der ISS wird durch Solarenergie über Sonnensegel gewährleistet. Die Solarzellen wandeln die Strahlungsenergie der Sonne in elektrische Energie um. Entlang der Gerüststruktur der ISS sind 16 Sonnenkollektoren aus jeweils 16 400 Solarzellen angebracht (Garcia 2017). Jeder Kollektor ist ca. 33,9 m lang und etwa 4,7 m breit (McDonald 2010). Je zwei Kollektoren sind zu einem Flügel verbaut, die entsprechend dem aktuellen Sonnenstand ausgerichtet werden (Abb. 5).

Die Solarkonstante gibt an, wie viel Strahlungsleistung auf die Solarzellen fällt und maximal für die Stromerzeugung zur Verfügung steht. Allerdings ist keine Solarzelle in der Lage, die auftreffende Strahlung vollständig in eine elektrische Leistung umzusetzen. Die Verluste sind teilweise erheblich. Das liegt auch daran, dass nur ein kleiner Teil des Lichtspektrums verwertet werden kann. Der

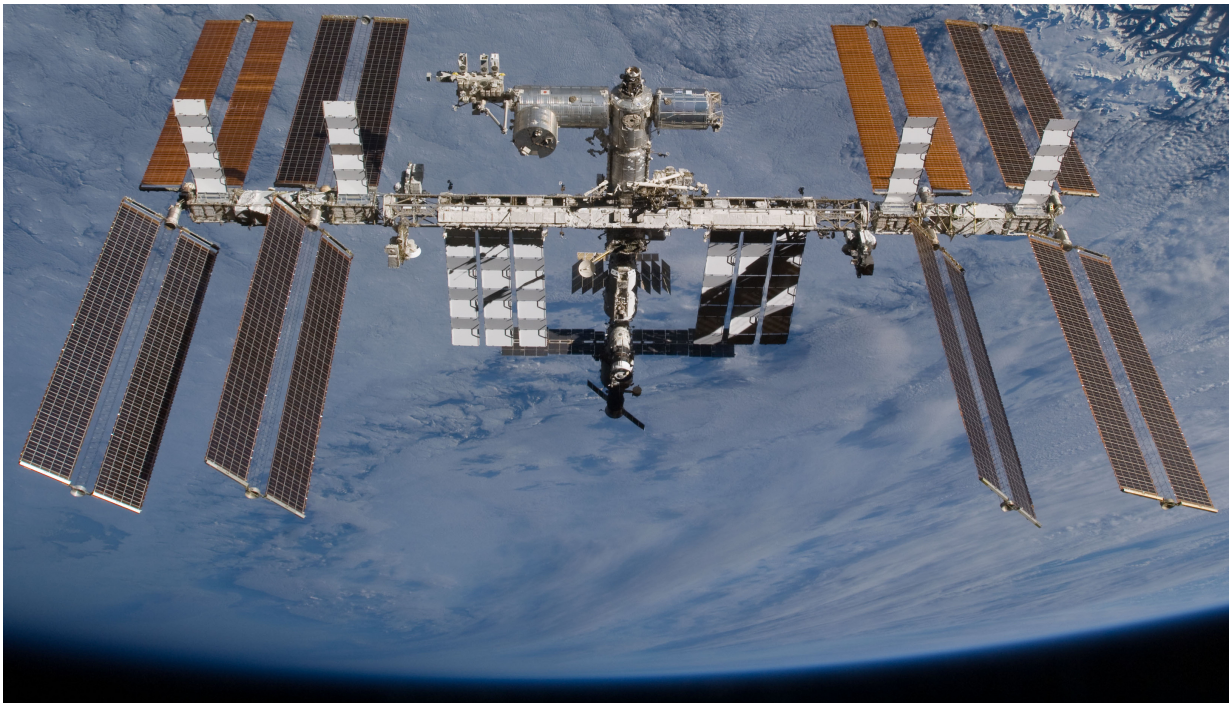

**Abbildung 5:** Foto der ISS aus dem Jahr 2009. Man erkennt die Solarmodule, die jeweils aus zwei einzelnen Kollektorflächen bestehen. Die Module drehen sich um ihre Längsachse um sich nach dem Sonnenstand auszurichten (Bild: NASA).

Wirkungsgrad $\eta$ gibt an, welcher Prozentsatz der eintreffenden Strahlungsleistung tatsächlich in Elektrizität umgewandelt werden kann. Für die Kollektoren der ISS gilt ein Wert von 14,5 %. Weitere Werte sind in Tab. 1 aufgeführt.

**Tabelle 1:** Wirkungsgrade verschiedener Typen von Solarzellen (Dimroth u. a. 2014; Quaschning 2013)

| Zellmaterial | Maximaler Wirkungsgrad im Labor | Maximaler Wirkungsgrad (Serienproduktion) | Typischer Modulwirkungsgrad |
|---|---|---|---|
| Monokristallines Silizium | 25,0 % | 22,9 % | 16 % |
| Polykristallines Silizium | 20,4 % | 17,8 % | 15 % |
| Amorphes Silizium | 12,5 % | 7,6 % | 6 % |
| CIS/CIGS[a] | 20,4 % | 15,1 % | 12 % |
| CdTe | 18,7 % | 12,8 % | 11 % |
| Konzentratorzelle | 43,6 – 44,7 % | 40,0 % | 30 % |

[a] Kupfer-Indium-Gallium-Diselenid bzw. Kupfer-Indium-Disulfid.

Während ihres Orbits um die Erde befindet sich die ISS zeitweise im Erdschatten. Dann empfangen die Solarzellen kein Sonnenlicht, um daraus elektrischen Strom zu produzieren. Stattdessen liefern Batterien die notwendige Versorgung mit Elektrizität. Sie werden durch die Solarmodule geladen, während die ISS von der Sonne beleuchtet wird (Garcia 2017).

## Die SPARTAN-Konsole

Die Versorgung der ISS mit Elektrizität (EPS, Electrical Power System) wird über ein Computersystem überwacht, das SPARTAN (Station Power, Articulation and Thermal Control) genannt wird. Die Daten können über verschiedene Darstellungen, auch Konsolen genannt, übersichtlich dargestellt werden (Abb. 6).

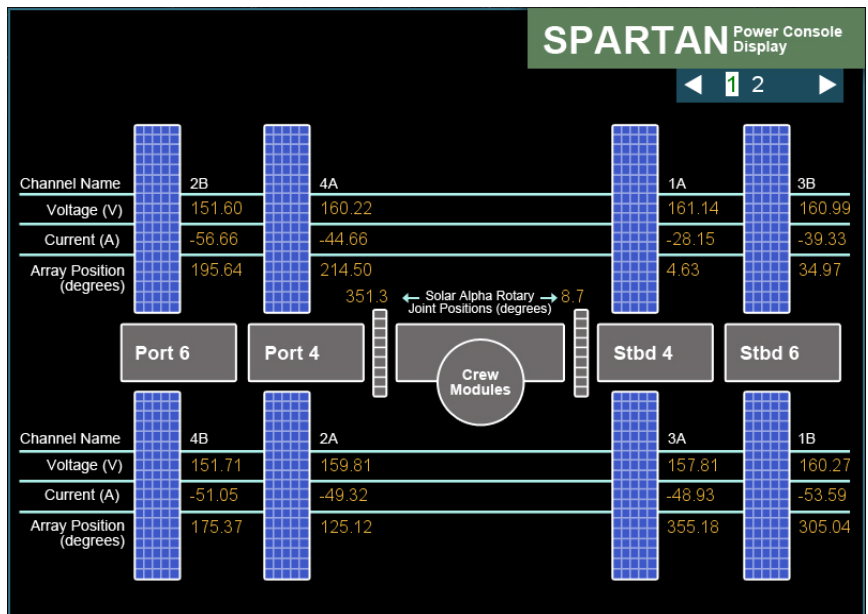

**Abbildung 6:** Die SPARTAN-Konsole dient zur Überwachung der elektrischen Versorgung der ISS (Bild: NASA, ISSLIVE).

Für jedes Solarmodul bestehend aus je zwei Kollektorflächen wird die aktuell erzeugte Spannung (Voltage) und Stromstärke (Current) dargestellt. Zudem werden die Rotationswinkel der Module zur Ausrichtung auf die Sonne aufgeführt.

## Aktivität: Solarstrom auf der ISS

### Vorbereitung für Lehrpersonen

Machen Sie sich mit der Internetseite `https://isslive.com` vertraut. Diese englischsprachige Seite vereint verschiedene Informationen rund um die ISS. Als Besonderheit ermöglicht sie einen einfachen Zugriff auf die Telemetrie der ISS und stellt sie in Form von übersichtlichen Konsolen dar. Wir werden für diese Aufgabe die Telemetrie des EPS nutzen, die unter der Rubrik SPARTAN erscheint. Bedenken Sie, dass auch die Schülerinnen und Schüler diese Seite für die Aufgabe nutzen werden, insofern Sie ausreichend Computer zur Verfügung stellen. Zur Not können auch Smartphones mit Internetzugang benutzt werden. Kann keine entsprechende Ausstattung gewährleistet werden, wird in der Aufgabe das Beispiel aus Abb. 6 und Tab. 2 benutzt.

Sie gelangen zur SPARTAN-Konsole (Abb. 6) über die Felder *LiveData* und anschließend *SPARTAN*. Die Konsole zeigt die Lage der acht Solarmodule, die mit je zwei Solarzellenflächen kombiniert sind. Sie sind wie in Tab. 2 bezeichnet. Für eine detailliertere Erläuterung der Funktionen können Sie dem Feld *Console Position Handbook* folgen. Die wesentlichen Informationen finden Sie aber auch in den Beschreibungen weiter oben.

Machen Sie sich mit den Aufgaben der Schülerinnen und Schüler vertraut. Fertigen Sie ausreichend Kopien der Arbeitsblätter an.

### Thematische Einführung (Vorschlag)

Diese Aktivität setzt voraus, dass Sie die Thematik zunächst grob einführen. Dies kann im Unterricht im Themenbereich der Grundbegriffe der Elektrizität geschehen. Sie können die Aufgabe motivieren, indem Sie den Schülerinnen und Schülern sagen, dass sie sich heute mit der Stromversorgung der Internationalen Raumstation beschäftigen werden. Fragen Sie zunächst, wer die ISS kennt. Vielleicht können manche der Kinder etwas über sie erzählen.

Sie können die Aufgabe mit einem Quiz einleiten. Fragen Sie die Schülerinnen und Schüler, wie die ISS mit Elektrizität versorgt wird. Sie können mehrere mögliche Antworten vorgeben, z. B. Brennstoffzellen, Solarzellen, Atomreaktoren, Windkraft. Falls Sie die Plattform *Kahoot!* kennen, nutzen Sie einfach folgenden Link:
`https://play.kahoot.it/#/k/4c9c9430-ebb1-4a0f-84c2-2e583ec8e6f1`

*Kahoot!* ist eine interaktive Online-Plattform für verschiedene didaktische Methoden. Typischerweise definiert die Lehrperson zunächst ein Quiz, das dann von den Schülerinnen und Schülern mit ihren Smartphones gespielt wird.
Sollten Sie *Kahoot!* nicht kennen oder nicht nutzen wollen, können Sie verschieden farbige Karten vorbereiten und den Farben die möglichen Antworten zuweisen. Das Spiel kann dann mit den Karten ebenso gut durchgeführt werden. Verraten Sie zunächst nicht die richtige Antwort.

Nutzen Sie das Resultat, um eine Diskussion über das korrekte Ergebnis unter den Schülerinnen und Schülern zu entfachen. Ermuntern Sie sie zu eigenen Recherchen, z. B. mit Wikipedia. Wiederholen Sie nach einigen Minuten das Quiz und lösen Sie dann das Rätsel auf. Sie können ergänzen, dass neben den Solarzellen auch Batterien verwendet werden, wenn die ISS nicht von der Sonne beleuchtet wird.

Lassen Sie die Schülerinnen und Schüler diskutieren, wofür auf der ISS Elektrizität benötigt wird.

Verteilen Sie dann die Arbeitsblätter und erläutern Sie die Aufgaben.

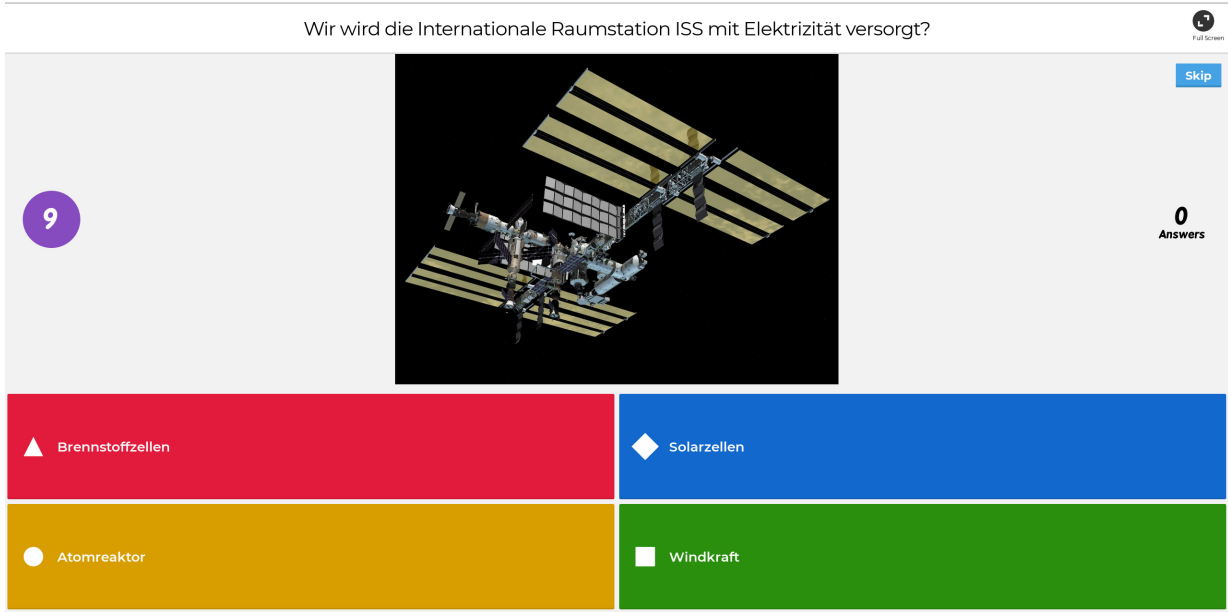

**Abbildung 7:** Mit *Kahoot!* lässt sich ein Quiz definieren, das die Schülerinnen und Schüler mit ihre Smartphones aktiv werden lässt.

## Aufgaben

Die Aufgabe besteht aus folgenden Teilschritten:

- Berechnung der Kollektorfläche der ISS
- Berechnung der maximal zur Stromerzeugung vorhandenen Strahlungsleistung der Sonne
- Ermittlung von Spannung und Stromstärke, die von den Solarzellen generiert werden
- Berechnung der zur Verfügung stehenden elektrischen Leistung
- Beantwortung von Fragen

### 1. Fläche der Solarzellen

Berechne aus den Angaben in den Hintergrundinformationen die Fläche eines der 16 Sonnenkollektoren und die Gesamtfläche aller Solarmodule. Wie berechnet man den Flächeninhalt eines Rechtecks?

### 2. Maximale Leistung

Wie hoch ist die pro Kollektor empfangene Strahlungsleistung, wenn man die Solarkonstante $E_0$ zugrunde legt? Der Wirkungsgrad $\eta$ gibt an, welcher Anteil der empfangenen Strahlungsleistung eine Solarzelle in elektrische Leistung umsetzen kann. Berechne unter der Annahme von $\eta = 0.145$ die maximal erreichbare elektrische Leistung für eines der 16 Module.

### 3a. Momentane Leistung (Online-Variante)

Die Stromversorgung durch die Solarzellen auf der ISS wird durch verschiedene System geregelt und überwacht. Des elektrische System (EPS, Electrical Power System) regelt die Versorgung. Das SPARTAN (Station Power, Articulation and Thermal Control) ermöglicht die Überwachung der wichtigsten Parameter.

Die englischsprachige Internetseite `https://isslive.com` enthält verschiedene Informationen rund um die ISS. Als Besonderheit ermöglicht sie einen einfachen Zugriff auf die Telemetrie der ISS und stellt sie in Form von übersichtlichen Konsolen dar. Die Telemetrie ist ein Datenstrom, der die technischen Parameter der ISS enthält. Wir werden für diese Aufgabe die Daten des EPS nutzen, die unter der Rubrik SPARTAN erscheint.

Wähle die Seite mit einem Computer, Tablet oder einem Smartphone an. Du gelangst zur SPARTAN-Konsole (Abb. 6) über die Felder *LiveData* und anschließend *SPARTAN*. Die Konsole zeigt die Lage der acht Solarmodule, die mit je zwei Solarzellenflächen kombiniert sind. Sie sind wie in Tab. 2 bezeichnet. In ähnlicher Weise kontrolliert ein Techniker der NASA im Kontrollzentrum den Zustand des Systems. Schau dir die Werte zur elektrischen Spannung (engl. Voltage) und Stromstärke (engl. Current) an und notiere Sie in Tab. 2. Berechne dann die elektrische Leistung $P$ für jedes Modul und die Summe daraus. Trage die Werte in der Tabelle ein.

**Tabelle 2:** Tabelle für die Spannungen und Stromstärken der acht Solarmodule.

| Modul | | $U$ (V) | $I$ (A) | $P$ (W) |
|---|---|---|---|---|
| 1 | A | | | |
| | B | | | |
| 2 | A | | | |
| | B | | | |
| 3 | A | | | |
| | B | | | |
| 4 | A | | | |
| | B | | | |
| | | | Gesamt | |

### 3b. Momentane Leistung (Offline-Variante)

Die Stromversorgung durch die Solarzellen auf der ISS wird durch verschiedene System geregelt und überwacht. In Abb.6 sind die entsprechenden Werte grafisch dargestellt. In ähnlicher Weise kontrolliert ein Techniker der NASA im Kontrollzentrum den Zustand des Systems. Schau dir die Werte zur elektrischen Spannung (engl. Voltage) und Stromstärke (engl. Current) an und notiere Sie in Tab. 2. Berechne dann die elektrische Leistung $P$ für jedes Modul und die Summe daraus. Trage die Werte in der Tabelle ein.

### 4. Auslastung

Die derzeit maximale elektrische Leistung, die die Solarzellen liefern, beträgt insgesamt 120 kW oder 120000 W. Berechne die momentane prozentuale Auslastung. Den aktuellen Wert entnimmst du Aufgabe 3.

### 5. Diskussion

Was könnten die Gründe sein, wenn die Solarzellen nicht unter maximaler Auslastung arbeiten?

Die ISS verbringt etwa ein Drittel der Zeit ihres 90minütigen Orbits im Erdschatten. Dort empfangen die Solarzellen kein Sonnenlicht. Woher bekommt die ISS dann den elektrischen Strom?

### Lösungen

#### 1. Fläche der Solarzellen

Aus den Hintergrundinformationen erfährt man, dass eines der 16 Kollektoren, von denen je zwei zu einem Modul zusammengeschlossen sind, 33,9 m lang und 4,7 m breit sind.

Den Flächeninhalt eines Rechtecks berechnet man durch Multiplizieren der langen mit der kurzen Seite.

$$A_{\text{Koll}} = \ell \cdot b = 33,9\,\text{m} \cdot 4,7\,\text{m} = 159,33\,\text{m}^2$$
$$A_{\text{tot}} = 16 \cdot A_{\text{Koll}} = 2549,28\,\text{m}^2$$

#### 2. Maximale Leistung

Die maximal verwertbare Strahlungsleistung kann durch Multiplizieren der Intensität der Sonne am Ort der Erde (Solarkonstante) mit der Kollektorfläche ermittelt werden.

$$P_{\text{abs}} = E_0 \cdot A_{\text{Koll}} = 1367\,\frac{\text{W}}{\text{m}^2} \cdot 159,33\,\text{m}^2 = 217804\,\text{W}$$

Um die daraus erzielbare elektrische Leistung zu bestimmen, wird die absorbierte Strahlungsleistung mit dem Wirkungsgrad $\eta$ multipliziert.

$$P_{\text{el, Koll}} = \eta \cdot P_{\text{abs}} = \eta \cdot E_0 \cdot A_{\text{Koll}} = 0,145 \cdot 1367\,\frac{\text{W}}{\text{m}^2} \cdot 159,33\,\text{m}^2 = 31582\,\text{W}$$

#### 3. Momentane Leistung

Die tatsächlichen Werte für Spannung und Stromstärke variieren zeitlich. Daher werden hier die Ergebnisse von der Offline-Variante gezeigt.

Die elektrische Leistung wird durch die Multiplikation von Spannung und Stromstärke berechnet.

$$P_{\text{el}} = U \cdot I$$

Die Ergebnisse sind in Tab. 3 aufgeführt.

#### 4. Auslastung

Die Auslastung ist das Verhältnis zwischen aktueller und maximal erreichbarer elektrischer Leistung, die die Solarmodule erzeugen.

$$\frac{P_{\text{akt}}}{P_{\text{max}}} = \frac{58555\,\text{W}}{120000\,\text{W}} = 0,488 = 48,8\%$$

In diesem Beispiel sind die Solarmodule etwa zur Hälfte ausgelastet.

Tabelle 3: Zusammenstellung der Spannungen und Stromstärken der acht Solarmodule.

| Modul | | $U$ (V) | $I$ (A) | $P = U \cdot I$ (W) |
|---|---|---|---|---|
| 1 | A | 161,14 | 28,15 | 4536 |
|   | B | 160,27 | 53,59 | 8589 |
| 2 | A | 159,81 | 49,32 | 7882 |
|   | B | 151,60 | 56,66 | 8590 |
| 3 | A | 157,81 | 48,93 | 7722 |
|   | B | 160,99 | 39,33 | 6332 |
| 4 | A | 160,22 | 44,66 | 7155 |
|   | B | 151,71 | 51,05 | 7745 |
| | | | Gesamt | 58555 |

**5. Diskussion**

Was könnten die Gründe sein, wenn die Solarzellen nicht unter maximaler Auslastung arbeiten?

- Ausrichtung der Solarmodule (angepasst an benötigter Versorgung)
- Beschädigung eines Moduls
- Alterung, Reduktion des Wirkungsgrads

Die ISS befindet sich zeitweise im Erdschatten. Dort empfangen die Solarzellen kein Sonnenlicht. Woher bekommt die ISS dann den elektrischen Strom?

- Batterien
- werden von den Solarzellen bei Beleuchtung geladen

## Literatur

## Danksagung

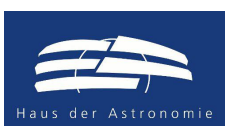
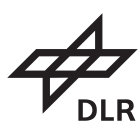
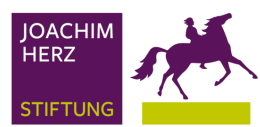